\documentclass[12pt]{article}
\usepackage{latexsym}
\begin{document}
\title{Sound Wave in Vortex with Sink}
\author{Soumen Basak\\
Physics Department\\   
The Institute of Mathematical Sciences\\
Tharamani\\
Chennai-600113\\
India}
\date{}
\maketitle
\begin{abstract}
  Using Komar's definition, we give expressions for the mass and
  angular momentum of a rotating acoustic black hole. We show that the
  mass and angular momentum so defined, obey the equilibrium version
  of the first law of Black Hole thermodynamics.  We also show that
  when a phonon passes by a vortex with a sink, its trajectory is
  bent. The angle of bending of the sound wave to leading order is
  quadratic in $A/cb$ and $B/cb$, where $b$ is the impact
  parameter and $A$ and $B$ are the parameters in the velocity of
  the fluid flow. The time delay in the propagation of sound wave
  which to first order depends only on $B/c^2$ and is independent of
  $A$.
\end{abstract}
\section{Introduction}
\label{Intro}
In 1981, Unruh\cite{bun,unruh} came up with the idea that the
phenomenon of Hawking radiation could be observed in an inviscid fluid
which has barotropic equation of state at sub-Planckian energy scales.
The idea for this model arose from a very simple question as to how a
sound wave propagates in an inhomogeneous fluid.  He showed that if
the fluid is barotropic and inviscid, and the flow of the fluid is
irrotational, the equation of motion that the fluctuation of the
velocity potential obeys is identical to that of a minimally coupled
massless scalar field in curved space-time Lorentzian geometry. Since
the fluid has barotropic equation of state, pressure force does not
generate vorticity. Thus an irrotational flow will remain
irrotational.The viscosity of the fluid is ignored in order to
maintain the Lorentzian character of the acoustic geometry.  The
fluid-flow will contain an analogue of a black hole if there is region
where it flows with supersonic speed. The boundary of this acoustic
black hole is a surface where the radial velocity of the fluid exceeds
the local speed of sound. It has also been shown by Unruh that an
acoustic black hole emits thermal radiation at a temperature given by
the surface gravity at the sonic horizon. In his demonstration he has
chosen a perfect sink flow $( \vec{v}=-A/r~\hat{r})$ of the fluid
which corresponds to non-rotating black hole at $~r= A/c~$ in the
acoustic geometry. Later this idea was extended by Matt Visser to a more
general situation where he considered a fluid motion which has both radial
and azimuthal components. This fluid flow corresponds to rotating
black hole at the same position in the acoustic geometry. This
correspondence has been borne out by the fact that such an acoustic
rotational black hole exhibits an analogue of superradiance as shown
by Basak and Majumdar~\cite{soumen, majumdar}.  As there is no
analogue of the Einstein equation that the fluid under consideration
obeys~\footnote{The fluid motion is governed by the Euler equation and
  the continuity equation}, it can mimic only the kinematical aspects
of gravity, not its dynamics.
\\
For asymptotically flat solutions of Einstein's equation one can
define the Komar charges corresponding to each isometry of the
spacetime. In the case of spacetime around a Kerr black hole Komar charges
corresponding to time translational and rotational Killing vectors are
called the mass and angular momentum of the black hole and they are
related by the  equilibrium version of the first law of black hole
thermodynamics. The basic criteria, for defining Komar charges and
also for those charges to obey equilibrium version of the first law of
black hole thermodynamics, have been discussed in
section(3). Interestingly acoustic geometry can also satisfy all those
criteria. Hence  we can define conserved charges for acoustic black hole
. Since the equilibrium version of the first law of black hole
thermodynamics does not depend on Einstein's equation, Komar charges
for the acoustic black hole will also satisfy this law.

One of the remarkable predictions of general relativity is that light
trajectories are bent in passing by a massive object because the
spacetime in the neighborhood of that massive object is curved. The
bending of light in the neighborhood of a massive object can be
described to sufficient accuracy using kinematical considerations
alone since the effect of the light on the geometry is negligible.
Moreover the source of curvature is not directly relevant to the
discussion; the light bending effect is a direct consequence of the
curved geometry which is locally Lorentzian. Since acoustic geometry
is an example of Lorentzian geometry, it is expected that this
phenomenon will also occur in this geometry if we consider a curved
geometry arising from a suitable choice of velocity of the fluid flow.

Based on the idea that sound waves experience an effective
gravitational field when passing through an inhomogeneous
vorticity-free flow of a barotropic inviscid fluid,
Fischer and Visser~\cite{Fischer} showed that when a sound wave passes by an
irrotational vortex, its trajectory is bent.  This is analogous to the light
bending effect in a physical gravitational field. In this paper we
have demonstrated, using differential geometric techniques which we
normally use in General Relativity, the bending of a sound wave for a
spiral flow (~vortex with sink~) of the fluid. We have assumed that
 the background density of the fluid is constant, which implies that
background pressure and the speed of sound are also constant and we
have treated the sound wave as a ray.  So our analysis is valid when$~
\frac{\lambda}{2~\pi}~$ is very  less compared to the  typical scale of
variation of the acoustic geometry, but it should be greater than the
interatomic distance because we have considered the fluid as a
continuous medium.

The plan of the paper is as follows. We first briefly discuss in
section~(\ref{acous}), the geometry of an acoustic black hole.  Using
the geometric properties of acoustic space-time, we define the mass
and angular momentum of an acoustic black hole in
section~(\ref{mass}). In section ~(\ref{eikonal}) we discuss the
eikonal approximation which we will use to study the propagation of a
sound wave in this geometry. In section~(\ref{sec:null}) we present an
analysis of null geodesics in acoustic geometry. This is followed by a
demonstration of the bending of a phonon in acoustic geometry and the
corresponding time delay in section~(\ref{bend}) and~(\ref{delay})
respectively. Experimental aspects have been discussed in
section~(\ref{outlook}).  Finally we conclude with a summary of the
results obtained in this paper.
\section{Acoustic Geometry}
\label{acous}
We consider the draining bathtub type of fluid motion with a line of
sink at the center, which is basically a (2+1) dimensional flow with a
sink at the origin.  This leads to the velocity
profile\cite{matt},\cite{vis} and \cite{visser} (in cylindrical polar
coordinates),
\begin{equation}
\overrightarrow{v}=-\frac{A}{r}~\hat{r}+\frac{B}{r}~\hat{\phi}~,
\label{vel}
\end{equation}
where$~A~$ and $~B~$ are real and positive .\\
A two surface, at$~r=\frac{A}{c}~$ in this flow, where the fluid
velocity is everywhere inward pointing and the radial component of the
fluid velocity exceeds the local sound velocity everywhere, behaves as
an outer trapped surface in this acoustic geometry.  This surface can
be identified with the future event horizon of the black hole. Since
the fluid velocity (\ref{vel}) is always inward pointing, the
linearized fluctuations originated in
the region bounded by the sonic horizon cannot cross this boundary .\\

A linearized fluctuation in the dynamical quantities (density and
pressure of the fluid, and velocity potential of the fluid motion)
leads to the fluctuations of velocity potential and the equation of
motion it satisfies is identical to that of a minimally coupled
massless scalar field ,
\begin{equation}
\frac{1}{\sqrt{-g}}~\partial_{\mu}\left(\sqrt{-g}~
g^{\mu\nu}~\partial_{\nu}\right)\Psi=0 
\label{wave}
\end{equation}
where$~g_{\mu\nu}~$ is acoustic metric and it is given by,
\begin{eqnarray}
{ds}^2=\left(\frac{\rho_{0}}{c}\right)^{2}\left[-\left(c^2-\frac{A^2+B^2}{r^2}\right){dt}^2+
\frac{2~A}{r}~dr~dt-2B~d\phi~dt+{dr}^2 +r^2~{d\phi}^2\right]
\label{metric}
\end{eqnarray}
If we assume that the background density of the fluid is constant, it
automatically implies that background pressure and the local speed of
sound are also constant. Thus we can ignore the position independent
pre-factor in the metric because it will not effect the equation of
motion of fluctuations of the velocity velocity potential.  As for the
Kerr black hole in general relativity, the radius of the boundary of
ergosphere of an acoustic black
hole is given by vanishing of $~g_{00}$, i.e,~$r_{ergo}=\sqrt{A^{2}+B^{2}}/c$.\\
\section{Mass and Angular Momentum of Acoustic Black Hole}
\label{mass}
In this section I will discuss the  definitions of certain conserved
quantities in the case of a acoustic black hole which can be thought
of as its mass and angular momentum. This is identical to Komar's
definition of mass and angular momentum using isometries of spacetime.
In his work, almost 40 years ago, Komar\cite{Komar} showed that for
every isometry $(\xi)$ of space time there exists a charge conserved$(Q_\xi)$ on
spatial hyper-surfaces$(\Sigma)$ which is given by,
\begin{equation}
Q_\xi=\int_{\Sigma}{}^{*}d\xi
\end{equation}
This conserved charge is arbitrary up to a constant factor, which can
be fixed using a known solution to the Einstein equation. For a Kerr
black hole the conserved charges corresponding to the time
translational and rotational Killing vectors are the mass$(M)$ and
angular momentum$(J)$ of the black hole and they are given by,
\begin{equation}
M_{H}=-\frac{1}{8\pi~G}\int_{H}{}^{*}dk~;
~~~~~~~~~J_{H}=\frac{1}{16\pi~G}\int_{H}{}^{*}dm
\label{mass-momentum}
\end{equation}
where $k$ and $m$ are time translational and rotational Killing vectors respectively. The signs in equation(\ref{mass-momentum}) reflect the signature of the spacetime geometry.\\
Using this result Smarr\cite{smarr} showed that if a stationary
axisymmetric spacetime contains a black hole then the mass and angular
momentum of the black hole with respect to an observer at the horizon
and one at asymptotic infinity are related as,
\\
\begin{equation}
M=M_{H}-\frac{1}{4\pi~G}~\int_{\Sigma}{}^{*}R(k)
\end{equation}
\begin{equation}
J=J_{H}+\frac{1}{8\pi~G}~\int_{\Sigma}{}^{*}R(m)
\end{equation}
\\
where ${}^{*}R(k)$ is Hodge dual of $R(k)(=R_{\mu\nu}k^{\nu}dx^{\mu}$).  $ M
$ and $ J $ are mass and angular momentum of the black hole with
respect to a stationary observer at infinity, and $M_{H}$ and $
J_{H}$ are the mass and angular momentum of the black hole with
respect to an observer at horizon. Since for vacuum solutions of the
Einstein's equation, $R(k)$ and $R(m)$ vanish, the mass and angular
momentum of a black hole with respect to observers at the horizon and
at asymptotic infinity are the same. If the spacetime contains a
Killing horizon such that the normal to the horizon is a linear
combination of time translational Killing vector and rotational
Killing vector as,
\begin{equation}
\ell=k~+~\Omega_{H}~m
\label{kv}
\end{equation}
where$~\Omega_{H}~$ is the angular velocity of the horizon, then the mass
at the horizon is given by\cite{wald},\cite{fro},
\begin{equation}
M_{H}=\frac{K_{H}~A_{H}}{4~\pi~G}~+~2~\Omega_{H}~J_{H}
\label{bhm}
\end{equation}
If the black hole has electric charge then the above equation is
modified to,
\begin{equation}
M=\frac{K_{H}~A_{H}}{4~\pi~G}~+~2~\Omega_{H}~J_{H}~+~V_{H}~Q_{H}+....
\label{bhm1}
\end{equation}
where $~V_{H}~$ is the electric potential at the horizon and $~
Q_{H}~$ is the charge of the black hole. If we restore all the occurrences of c, equation(\ref{bhm}) is modified to the following form,
\begin{equation}
M_{H}=\frac{K_{H}~A_{H}}{4~\pi~G}~+~\frac{2~\Omega_{H}~J_{H}}{c^{2}}
\label{bhm2}
\end{equation}
Even though in order to derive the equation(\ref{bhm1}) for charged black hole,
one has to use Einstein's equation with energy-momentum tensor of
electromagnetic fields, equation(\ref{bhm2}) is insensitive to the
Einstein's equation.  This equation entirely depends upon the
geometric properties of the spacetime and does not depend upon how
geometry of the spacetime is changing with time.  The above equation
will hold if a curved spacetime has the following properties,
\begin{itemize}
\item
Geometry of the spacetime is pseudo-Riemannian  with  signature
(-,+,+,+)
\item
Spacetime is asymptotically flat which implies that with increasing r ,
conserved current,$~J^{\mu}=-R^{\mu}_{~\nu}\xi^{\nu}$ decreases faster
than
the increase of area of the constant radius surface.
\item
Killing vectors of the spacetime geometry are time-translational
and rotational.
\item
A Killing horizon is present with outward normal
$~l=k~+~\Omega_{H}~m~$
\end{itemize}

Acoustic geometry corresponding to the metric(\ref{metric}) is similar
to the geometry of a curved spacetime containing a rotating black hole
as far as kinematics are concerned.  There are two isometries in this
geometry and the corresponding Killing vectors
are $k^{\mu}=\delta^{\mu}_{t}~$ and $~m^{\mu}=\delta^{\mu}_{\phi}$.
The normal to the horizon is $l^{\mu}=k^{\mu}+\Omega_{H}m^{\mu}$.
Since it is a linear combination of two Killing vectors and
$\Omega_{H}=\frac{B~c^{2}}{A^{2}}$ is constant, this vector is also
a Killing vector of the acoustic metric. Therefore the event horizon
of the acoustic black hole is a Killing horizon. So we can define the
surface gravity of that Killing horizon as,
\begin{equation}
l^{\mu}\nabla_{\mu}l_{\nu}=\frac{K_{H}}{c} l_{\nu}
\end{equation}
where$~ K_{H}~$is the surface gravity of the horizon which is equal to
 $ c^{3}/A$.  Hence this geometry is  not only Lorentzian, but also
satisfies all the four properties mentioned above.  Therefore we can
define the mass and angular momentum of the rotating acoustic black hole
as the conserved charges corresponding to the time translational and
rotational Killing vectors of the acoustic geometry. 
According to this, the conserved charges at the horizon corresponding to
the time translational Killing vector is given by,
\begin{eqnarray}
Q_{k}&=&-\frac{c}{8~\pi}\int_{H}{}^{*}dk\nonumber\\
&=&-\frac{c}{8~\pi}\int_{H}~\partial_{\mu}k_{\nu}~
\epsilon^{~\mu\nu}_{~~~~\alpha}~dx^{\alpha}
\nonumber\\
\nonumber\\
&=&\frac{(A^{2}+B^{2})~c^{2}}{2~A^{2}}
\end{eqnarray}
\\
and the conserved charge at the horizon corresponding to rotational
Killing vector is given by,
\begin{eqnarray}
Q_{m}&=&\frac{c^{3}}{16~\pi}\int_{H}{}^{*}dm\nonumber\\
&=&\frac{c^{3}}{16~\pi}\int_{H}~\partial_{\mu}m_{\nu}~\epsilon^{~\mu\nu}
_{~~~~\alpha}~dx^{\alpha}\nonumber\\
\nonumber\\
&=&\frac{B~c^{2}}{4}
\end{eqnarray}
These two quantities satisfy the following relation ,
\begin{equation}
Q_{k}=\frac{K_{H}~L_{H}}{4~\pi}~+~\frac{2~\Omega_{H}~Q_{m}}{c^{2}}
\end{equation}
Here $L_{H}(=2\pi\frac{A}{c})$ is the length of the horizon.\\
This equation is identical to the relation between mass and angular momentum
of physical rotating black hole in $G=1$ unit.\\
\\
The most interesting feature of these results is that the angular
momentum is proportional to the velocity parameter B. If one switches
off the rotational motion of the fluid, B will become zero and hence the
angular momentum of the black hole disappears. This is expected because
in the absence of B, the horizon is no longer rotating. If the vortex
motion is quantized, i.e, if B takes values proportional to an
integer, then $Q_{m}$ and $Q_{k}$ take discrete values.  From the
expression of $Q_{m}$ and $Q_{k}$ it is clear that $Q_{m}$ is
proportional to integer and $Q_{k}$ has a minimum value $c^{2}/2$
. So we can say that this acoustic black hole is analogous to a physical
black hole whose mass and angular momentum are quantized. Since there
is no analog of Einstein's equation in acoustic analog models of
gravity, the  relation between the variation of $Q_{k}$ and $Q_{m}$ is not
identical to that of physical process version of first law of black
hole thermodynamics. 
\section{Eikonal Approximation} 
\label{eikonal}
We now analyze the geometric properties of acoustic space-time in
terms of the null geodesics. We will consider the propagation of a
sound wave in this geometry as a ray. This is valid in the
eikonal approximation \cite{wald},\cite{mis}which we describe briefly.
\\
The eikonal approximation is described by the two conditions,
\begin{itemize}
\item
Reduced wavelength$(\frac{\lambda}{2\pi})$ of the sound wave is
very small compared to the typical radius of curvature of the acoustic
geometry.
\item
Reduced wavelength$(\frac{\lambda}{2\pi})$ of the sound wave is
very small compared to typical length over which the amplitude,
polarization and wavelength of the wave change.
\end{itemize}
Under these conditions we can split the  scalar field $\Psi$ into a rapidly
changing real phase and a slowly changing complex amplitude,
\begin{eqnarray}
\Psi(x)&=&Re\left\{amplitude \times \exp\left(\frac{i}{\epsilon}
~\theta(x)\right) \right\}\nonumber\\
&=&Re\left\{(a+ \epsilon~b+\epsilon^{2}~c + ......  ) \times 
\exp\left(\frac{i}{\epsilon}~\theta(x) \right) \right\}
\end{eqnarray} 
Here $\epsilon$ is a very small parameter. If$~R~$is the typical radius
of curvature of the  acoustic geometry and $L $ is the typical length over
which the amplitude and the polarization of the wave remains unchanged,
then the order of magnitude of  $\epsilon$ is
$~\frac{\lambda}{2\pi}\Big{/}Min(R,L)~$. The first
term $(~a~)$ in the amplitude is wave length independent. It is the 
dominant term in the expression while the rest of the terms are wave
length dependent and also very small compared to the first term.
Mathematically, the scalar field $(~\Psi~)$ corresponds to a massless
particle in the acoustic geometry when its frequency tends to
infinity,i.e, ($\lambda / 2\pi\longrightarrow 0$ ).  Practically all
the particles have finite de Broglie wave length. This causes the
wavelength dependent corrections to the amplitude of the scalar
field. These correction terms disappear in the $\lambda /
2\pi\longrightarrow 0$ limit.

Substituting this in equation(\ref{metric}) and then collecting terms
of order $\frac{1}{\epsilon^{2}} $ and $\frac{1}{\epsilon} $ we get,
\begin{equation}
\nabla_{\mu}\theta~\nabla^{\mu}\theta=0
\end{equation}
and,
\begin{equation}
\nabla^{\mu}\theta~\nabla_{\mu}a = -\frac{1}{2}(\nabla_{\mu}
\nabla^{\mu}\theta)~a 
\end{equation}
Since sound rays are defined as the curves normal to the constant
phase$~\theta(x)~$ surface and the wave vector
$k_{\mu}=\nabla_{\mu}{\theta}$ is normal to these surfaces, the
trajectory of a sound ray is described by integral curves of wave
vector.

According to this definition of wave vector(k),
\begin{equation}
k_{\mu}k^{\mu}=0
\label{null}
\end{equation}
i.e, trajectories of sound ray are null curves and 
\begin{equation}
k^{\mu}~\nabla_{\mu}a = -\frac{1}{2}(\nabla_{\mu}
k^{\mu})~a 
\label{propagation}
\end{equation}
Since the wave vector is the gradient of a scalar function, equation
(\ref{null}) implies that the integral curves of the wave vector are also
null geodesics i.e,
\begin{equation}
k^{\mu}\nabla_{\mu}k^{\nu}=0
\label{geodesic}
\end{equation}

Equation(\ref{propagation}) leads to two important results,

(1) If we define polarization of sound wave as,
\begin{equation}
f=\frac{a}{\sqrt{a\overline{a}}}
\end{equation}
then,
\begin{equation}
f~\overline{f}=1
\end{equation}
and,
\begin{equation}
k^{\mu}\nabla_{\mu}f=0,
\end{equation}
So the polarization f is parallel transported along the integral curve  of
$k^{\mu}$.

(2)We can define a conserved vector$~j_{\mu}=
a\overline{a}k_{\mu}~$,satisfying\begin{equation}
  \nabla_{\mu}j^{\mu}=0
\end{equation}
The corresponding conserved charge on spacelike hypersurfaces of
acoustic geometry is,
\begin{equation}
Q=\int_{\Sigma}~j^{\mu}~d^{3}\Sigma_{\mu}        
\end{equation}
Physically this conservation corresponds to conservation of sound
rays ( phonon number).  Therefore in the eikonal approximation, we can
treat the sound wave as a ray (phonon) which is a particle of zero rest
mass with four momentum $p=\hbar k$ moving along  null geodesics and
parallel transporting the polarization f. Hence we will consider a phonon
instead of sound wave in the analysis of null geodesics in acoustic
geometry.

The curvature (Ricci Scalar) of the acoustic geometry is,
$~R=2~r_{ergo}^{2}/r^{4}~$.  Our analysis is valid if $\lambda\over
2\pi$~$\ll$~$ r^{2}/\sqrt{2}r_{ergo}$. Physically, when the wavelength
of the phonon is less than the scale of variation of the metric, the
effect of the curvature on the phonon's trajectory is significant.
\section{Null Geodesics in Acoustic Geometry}
\label{sec:null}

As mentioned earlier there are two isometries $(\xi_{A})$ of the acoustic
geometry. Corresponding to these we have the quantities conserved
along the null geodesics given by,
\begin{eqnarray}
K_{A}=-(\xi_{A})^\mu ~g_{\mu\nu}~\frac{dx^{\nu}}{d\lambda} 
\label{conserved quantity}
\end{eqnarray}
Here $~\lambda~$ is the affine parameter of the geodesics and $~ A~$
takes values 1 and 2.  Now using the components of the acoustic
metric in the above equation we get the following equations,
\begin{equation}
K_{1}=\left(c^{2}-\frac{A^{2}+B^{2}}{r^{2}}\right)~\frac{dt}
{d\lambda}-\frac{A}{r}~\frac{dr}{d\lambda}+B~\frac{d\phi}{d\lambda}
\label{Conserved1}
\end{equation}
\\
\begin{equation}
K_{2}=B~\frac{dt}{d\lambda}-r^{2}~\frac{d\phi}{d\lambda}
\label{Conserved2}
\end{equation}
\\
In order to find the expression for $\frac{dr}{d\lambda}$ 
we have made the following substitutions in equations,
\begin{equation}
dt=dT+\frac{A~r}{r^{2}~c^{2}-A^{2}}~dr
\label{T}
\end{equation}
\\
\begin{equation}
d\phi=d\chi+\frac{A~B}{r(r^{2}~c^{2}-A^{2})}~dr
\label{chi}
\end{equation}
\\
Due to these transformations, equations~(\ref{Conserved1})and(\ref{Conserved2}) are modified to following equations,
\begin{equation}
K_{1}=\left(c^{2}-\frac{A^{2}+B^{2}}{r^{2}}\right)
~\frac{dT}{d\lambda} + B~\frac{d\chi}{d\lambda} 
\label{Conservedt}
\end{equation}
\nonumber\\
\begin{equation}
K_{2}=B~\frac{dT}{d\lambda} - r^{2}~\frac{d\chi}
{d\lambda}
\label{Conservedphi}
\end{equation}
\nonumber\\
Solving equations
(\ref{Conservedt})~and~(\ref{Conservedphi})~we get,
\begin{equation}
\frac{dT}{d\lambda}=\frac{K_{1}~r^{2}~+K_{2}~B}
{(r^{2}~c^{2}-A^{2})}
\end{equation}
\\
\begin{equation}
\frac{d\chi}{d\lambda}=\frac{K_{1}~B~r^{2}~
+K_{2}~(A^{2}~+~B^{2}~-~r^{2}c^{2})}{r^{2}(r^{2}~c^{2}-A^{2})}
\end{equation}
\\
Since$~\lambda~$ is the affine parameter of the null geodesics in
acoustic geometry, the trajectory of a phonon can be determined by the
following equation,
\begin{equation}
g_{\mu\nu}~\frac{dx^{\mu}}{d\lambda}~\frac{dx^{\nu}}{d\lambda}=0
\end{equation}
\\
Using the components of the acoustic metric in the above equation, we get,

$$\frac{r^{2}~c^{2}}{r^{2}~c^{2}-A^{2}}~\left(\frac{dr}{d\lambda}\right)^{2}
-\left(c^{2}-\frac{A^{2}+B^{2}}{r^{2}}\right)~\left(\frac{dT}{d\lambda}
\right)^{2}-2~B~\frac{dT}{d\lambda}~\frac{d\chi}{d\lambda}~$$
\begin{equation}
+~r^{2}~\left(\frac{dT}{d\lambda}\right)^{2}~
+~\left(\frac{dz}{d\lambda}\right)^{2}=0
\label{nullgeo}
\end{equation}
\\
Substituting $~\frac{dT}{d\lambda}~$ and$~\frac{d\chi}{d\lambda}~$
in  equation(\ref{nullgeo}) we get,
\\
\begin{equation}
\left(\frac{dr}{d\lambda}\right)^{2}=\frac{K_{1}^{2}}{c^{2}}~+
~\frac{2~K_{1}K_{2}~B}{r^{2}~c^{2}}
-\frac{K_{2}^{2}}{r^{2}}\left(1-\frac{A^{2}+B^{2}}{r^{2}~c^{2}}\right)
\end{equation}
\\
Rearranging the above equation we get,
\begin{equation}
\frac{1}{2}\left(\frac{dr}{d\lambda}\right)^{2}~+~V(r)
=\frac{1}{2}\frac{K_{1}^{2}}{c^{2}}
\label{potential}
\end{equation}
where,
\begin{equation}
V(r)=\frac{K_{2}^{2}}{2~r^{2}}\left(1-\frac{A^{2}+B^{2}}
{r^{2}~c^{2}}\right)~-~\frac{K_{1}K_{2}~B}{r^{2}~c^{2}}
\label{pot}
\end{equation}
\\ This equation describes the radial motion of a sound ray in
acoustic geometry and it is identical to the  one dimensional
non-relativistic motion of a particle of unit mass and energy
$~\frac{K_{1}^{2}}{2~c^{2}}~$ in a potential$~V(r)~$\cite{wald}.
\\
The extrema of this potential, where$~\frac{dV(r)}{dr}=0$, are at,
\\
\begin{equation}
r=\sqrt{\frac{2~K_{2}~(A^{2}+B^{2})}{~K_{2}~c^{2}-2~B~K_{1}}}
\end{equation}
\\
 There is a difference in the nature of the effective
potential $V(r)$ for $K_{2} < 0$ as compared with for  $K_{2}$.
 If $K_{2} < 0~$ is negative then this potential has
maxima at,
\\
\begin{equation}
r_{M}=\sqrt{\frac{2~K_{2}~(A^{2}+B^{2})}{~K_{2}~c^{2}-2~B~K_{1}}}
\label{rmax}
\end{equation}
\\
for all negative values of $K_{2}$ because the 2nd derivative of this
potential with respect to $r$ is negative at $~r=r_{M}~$ for all
negative values of $K_{2}$. But when $~K_{2}> 0~$, $V(r)$ does not
have any extrema (i.e, $\frac{dV(r)}{dr}=0~$, does not have any real
root in $r$) if $K_{2}~c^{2} \le 2~B~K_{1}$. Therefore when $K_{2}
~$ is positive, for the existence of a maximum in $V(r)$, we must
have $K_{2}~c^{2} >2~B~K_{1}$. If a sound wave moves towards the
center of the vortex such that $V(r_{M})>
\frac{1}{2}\frac{{K_{1}}^{2}}{c^{2}}$, it will be reflected at the
turning point where $\frac{dr}{d\lambda}=0~$. Considering the positive
solutions of $r$ of equation(\ref{potential}), we get 
position of the largest turning point of sound ray  at,
\\
\begin{equation}
r_{T}=\sqrt{~\frac{K_{2}~(~K_{2}~c^{2}-2~B~K_{1})}{2{K_{1}}^{2}}
\left(1~+~\sqrt{1-\frac{4~{K_{1}}^{2}~(A^{2}+B^{2})~}
{(K_{2}~c^{2}-2~B~K_{1})^{2}}}\right)}
\label{rturn}
\end{equation}
\\
If  $~b(=|K_{2}|~c/K_{1})$ is the magnitude of the impact parameter
and $~(A,B)\ll~c~b~$ then expanding $r_{T}$ in series~(up to 2nd
order)~ in the small quantities $\frac{A}{c~b}$ and $\frac{B}{c~b}$ we
get,
\\
\begin{equation}
r_{T}=b~\left[~1~-~\frac{K_{2}}{|K_{2}|}~\frac{B}{c~b}
~-~\frac{1}{2}~\frac{A^{2}}{c^{2}~b^{2}}~-~\frac{B^{2}}{c^{2}~b^{2}}~
+~\mathcal{O}\left\{\left(\frac{A}{cb}\right)^{3},
\left(\frac{B}{cb}\right)^{3}\right\}\right]
\end{equation}
\\
This shows that, when $K_{2}$ is negative (i.e angular momentum of
the sound ray is positive), $r_{T}$ is greater than impact
parameter,$~b~$. Therefore in order to make the distance of closest
approach shorter than the impact parameter we have to consider
positive values of $~K_{2}~$. Since$~\frac{dr}{d\lambda}~$ takes both
positive and negative value, to find the bending angle we have to
choose the sign of $~\frac{dr}{d\lambda~}$. The choice is made on the
basis of fact that when $A=B=0$, we should obtain the Newtonian
result, i.e, duration of propagation sound wave through the fluid
should be positive.
\\

\section{Bending of Phonon Trajectory}
\label{bend}
In this section we will calculate the angle of bending of a phonon
ray, expanding around $A=0~$ and $B=0$, as it passes by irrotational
vortex. Before calculating the angle of bending one  important thing
has to be noticed. Since  $d\phi\over dr$ is not symmetric
about the turning point even if $dr\over d\lambda$ flips sign at that
point, the  change in the polar angle when the phonon travels
towards the vortex is not same as that when phonon travels away from
the vortex.  Therefore the total change of the polar angle throughout
its path is given by,
\\
\begin{eqnarray}
\Delta\phi&=&\int_{\infty}^{r_{T}}~{d\phi\over dr}~dr~
+~\int_{r_{T}}^{\infty}~{d\phi\over dr}~dr\nonumber\\
\nonumber\\
&=&\Delta\phi_{-}~+~\Delta\phi_{+}
\end{eqnarray}
\nonumber\\
Expanding $~\Delta\phi_{-}~$ around $~A=B=0~$ we get,
\begin{eqnarray}
\Delta\phi_{-}&=&(\Delta\phi_{-})_{A=B=0}~+~A~\left(\frac{d}{dA}(\Delta\phi_{-})\right)_{A=B=0}~+~B~\left(\frac{d}{dB}(\Delta\phi_{-})\right)_{A=B=0}\nonumber\\\nonumber\\
\nonumber\\
&+&~\frac{A^{2}}{2}~\left(\frac{d^{2}}{dA^{2}}(\Delta\phi_{-})\right)_{A=B=0}
+~\frac{B^{2}}{2}~\left(\frac{d^{2}}{dB^{2}}(\Delta\phi_{-})\right)_{A=B=0}\nonumber\\
\nonumber\\
\nonumber\\
&+&~A~B~\left(\frac{d^{2}}{dAdB}(\Delta\phi_{-})\right)_{A=B=0}
~+~\mathcal{O}\left\{\left(\frac{A}{cb}\right)^{3},
\left(\frac{B}{cb}\right)^{3}\right\}\nonumber\\
\nonumber\\
&=&-\mathrm{sgn}(K_{2})\left[\frac{\pi}{2}+~\frac{3~\pi}{8}
\frac{(A^{2}+B^{2})}{b^{2}~c^{2}}\right]~-~\frac{1}{2}\frac{A~B}{b^{2}~c^{2}}\nonumber\\
\nonumber\\
&+&{\mathcal{O}\left\{\left(\frac{A}{cb}\right)^{3},
\left(\frac{B}{cb}\right)^{3}\right\}}
\end{eqnarray}
Doing the same for $~\Delta\phi_{+}~$ we get,
\begin{eqnarray}
\Delta\phi_{+}&=&(\Delta\phi_{+})_{A=B=0}~+~A~\left(\frac{d}{dA}(\Delta\phi_{+})\right)_{A=B=0}~+~B~\left(\frac{d}{dB}(\Delta\phi_{+})\right)_{A=B=0}\nonumber\\\nonumber\\
\nonumber\\
&+&~\frac{A^{2}}{2}~\left(\frac{d^{2}}{dA^{2}}(\Delta\phi_{+})\right)_{A=B=0}
+~\frac{B^{2}}{2}~\left(\frac{d^{2}}{dB^{2}}(\Delta\phi_{+})\right)_{A=B=0}\nonumber\\
\nonumber\\
\nonumber\\
&+&~A~B~\left(\frac{d^{2}}{dAdB}(\Delta\phi_{+})\right)_{A=B=0}~
+~\mathcal{O}\left\{\left(\frac{A}{cb}\right)^{3},
\left(\frac{B}{cb}\right)^{3}\right\}\nonumber\\
\nonumber\\
&=&-\mathrm{sgn}(K_{2})\left[\frac{\pi}{2}+~\frac{3~\pi}{8}\frac{(A^{2}+B^{2})}{b^{2}~c^{2}}\right]~+~\frac{1}{2}\frac{A~B}{b^{2}~c^{2}}\nonumber\\
\nonumber\\
&+&\mathcal{O}\left\{\left(\frac{A}{cb}\right)^{3},
\left(\frac{B}{cb}\right)^{3}\right\}
\end{eqnarray}
Hence the total change of polar angle of a phonon trajectory is,
\begin{equation}
\Delta\phi=-\mathrm{sgn}(K_{2})\left[\pi +\left\{~\frac{3~\pi}{4}(A^{2}+B^{2})\right\}\frac{1}{b^{2}~c^{2}}\right]~+~\mathcal{O}\left\{\left(\frac{A}{cb}\right)^{3},
\left(\frac{B}{cb}\right)^{3}\right\}
\label{polarangle}
\end{equation}
\\
The first term corresponds to the case when fluid velocity
 $\vec{v}=0$.  This matches with the Newtonian result. We have
considered terms to leading order in $A$ and $B$, which in this
case is the 2nd order, but the terms corresponding to AB
for $\Delta\phi_{-}$ and $\Delta\phi_{+}$ cancel each other and
hence do not appear in the expression. Physically when the phonon
approaches the turning point its motion is aided by the radial fluid
motion, while it is hindered after crossing the turning point. The
actual angle of bending is given by,
\\
\begin{eqnarray}
\delta\phi&=&\Delta\phi~-~\Delta\phi~\Big{|}_{A=B=0}\nonumber\\
\nonumber\\
&=&-\mathrm{sgn}(K_{2})\left[~\frac{3~\pi}{4}(A^{2}+B^{2})\right]\frac{1}{b^{2}~c^{2}}+~\mathcal{O}\left[\left(\frac{A}{cb}\right)^{3},
\left(\frac{B}{cb}\right)^{3}\right]\nonumber\\
\nonumber\\
&=&-\mathrm{sgn}(K_{2})\frac{3~\pi}{4}\left(\frac{r_{ergo}}{b}\right)^{2}
+~\mathcal{O}\left[\left(\frac{A}{cb}\right)^{3},
\left(\frac{B}{cb}\right)^{3}\right]
\label{bendingangle}
\end{eqnarray}
This is consistent with Fischer and Visser's\cite{Fischer} result for
an irrotational vortex without a sink. Here the integrals are slightly
tricky because the radius of the turning point depends both on A and
B. So when we differentiate $\Delta\phi$ with respect to $A$ and $B$ ,
we take this into account. We have done the integration from $r_{T}$
to $\infty$. This is a good approximation because the radius of the
boundary of the vortex is very large compared to the turning point.

\section{Time delay and advance} 
\label{delay}
Unlike in the case of the bending of a phonon, in order to calculate the time
delay and time advance\cite{uwefischer,Alcub,Olum,mattvisser,Clark,Natario} in phonon propagation, we will compare null
geodesics of same turning point for the fluid with motion and the fluid
at rest. Using  equation (\ref{rturn}) we express $K_{2}$ in terms of $r_{T}$ ,
\begin{equation}
K_{2}=\frac{K_{1}r_{T}^{2}\left(B\pm \sqrt{c^{2}r_{T}^{2}-A^{2}}\right)}
{\left(c^{2}r_{T}^{2}-A^{2}-B^{2}\right)}
\label{K_{2}}
\end{equation}
Substituting $K_{2}$ in $\frac{dr}{d\lambda}$ and $\frac{dt}{d\lambda}$ we get, time interval in which a phonon passes through the vortex,
\begin{eqnarray}
\Delta t&=&\int_{r_{V}}^{r_{T}}~{dt\over dr}~dr~+
~\int_{r_{T}}^{r_{V}}~{dt\over dr}~dr\nonumber\\
\nonumber\\
&=&\Delta t_{-}~+~\Delta t_{+}
\end{eqnarray}
where $r_{V}$ is the radius of the vortex.
\\

Expanding $\Delta t_{-}$ around $A=B=0$ for fixed $r_{T}$ we get,
\begin{eqnarray}
\Delta t_{-}&=&(\Delta t_{-})_{A=B=0}~+~A~\left(\frac{d}{dA}(\Delta t_{-})\right)_{A=B=0}~+~B~\left(\frac{d}{dB}(\Delta t_{-})\right)_{A=B=0}\nonumber\\\nonumber\\
&~+~&\mathcal{O}\left\{\left(\frac{A}{cb}\right)^{2},
\left(\frac{B}{cb}\right)^{2}\right\}\nonumber\\
\nonumber\\
&=&\frac{1}{c}\sqrt{r_{V}^{2}-r_{T}^{2}}+\mathrm{sgn}(K_{2})~\frac{B}{c^{2}}\tan^{-1}\left(\sqrt{\left(\frac{r_{V}}{r_{T}}\right)^{2}-1}~\right)\nonumber\\
&-&\frac{A}{c^{2}}\ln\left(\frac{r_{V}}{r_{T}}\right)~+~\mathcal{O}\left\{\left(\frac{A}{c^{2}}\right)^{2},
\left(\frac{B}{c^{2}}\right)^{2}\right\}
\end{eqnarray}
\\
Doing the same for $\Delta t_{+}$ we get,
\begin{eqnarray}
\Delta t_{+}&=&(\Delta t_{+})_{A=B=0}~+~A~\left(\frac{d}{dA}(\Delta t_{+})\right)_{A=B=0}~+~B~\left(\frac{d}{dB}(\Delta t_{+})\right)_{A=B=0}\nonumber\\\nonumber\\
&~+~&\mathcal{O}\left\{\left(\frac{A}{cb}\right)^{2},
\left(\frac{B}{cb}\right)^{2}\right\}\nonumber\\
\nonumber\\
&=&\frac{1}{c}\sqrt{r_{V}^{2}-r_{T}^{2}}+\mathrm{sgn}(K_{2})~\frac{B}{c^{2}}\tan^{-1}\left(\sqrt{\left(\frac{r_{V}}{r_{T}}\right)^{2}-1}~\right)\nonumber\\
&+&\frac{A}{c^{2}}\ln\left(\frac{r_{V}}{r_{T}}\right)~+~\mathcal{O}\left\{\left(\frac{A}{c^{2}}\right)^{2},
\left(\frac{B}{c^{2}}\right)^{2}\right\}
\end{eqnarray}
\\
In contrast to the angle of bending, the time delay to leading order is linear in $A$ and $B$. If we assume that $( A/c^{2}\tau, B/c^{2}\tau)$
$ \ll 1 $,where $\tau = \frac{1}{c}\sqrt{r_{V}^{2}-r_{T}^{2}}$ ,we can neglect the higher order terms. Subtracting the
zeroth order term from$~\Delta t$ we get the time delay up to first
order in A and B to be,
\begin{eqnarray}
\delta t &=&\Delta t~-~\Delta t~\Big{|}_{A=B=0}\nonumber\\
\nonumber\\
&=&\mathrm{sgn}(K_{2})~\frac{2 B}{c^{2}}\tan^{-1}\left(\sqrt{\left(\frac{r_{V}}{r_{T}}\right)^{2}-1}~\right)+\mathcal{O}\left\{\left(\frac{A}{c^{2}}\right)^{2},\left(\frac{B}{c^{2}}\right)^{2}\right\}
\end{eqnarray}
\\
The only difference is in the limits of integration. This is because
of the fact that the zeroth order term blows up if we put infinity in
place of the radius of the vortex. In fact, the radius of the vortex,
in actual experiment is large but finite. So a phonon will pass
through the vortex in a finite time interval.  The effect of the
radial flow of the fluid on the propagation of the phonon is different
during the two stages of motion -- before and after having crossed the
turning point.  When the phonon approaches the turning point its
motion is aided by the radial fluid motion, while it is hindered
after crossing the turning point.  The effect of the rotational
velocity of the fluid is the same throughout the motion. Depending on
the sign of $K_2$ and the angular momentum of the fluid, the sign of
the time ``delay'' is decided by the combined effect of these. The
terms that are  first order in $A/c^{2}$ of $\Delta t_{+}$ and
$\Delta t_{-}$ cancel each other.  This is because of the fact that
when the radial coordinate is decreasing the radial flow of the fluid
favors its motion and in the opposite case it opposes its motion.

\section{Experimental Outlook}
\label{outlook}
Unlike Hawking radiation and Superradiance for gravitational
black-holes, the phenomena of bending of light and the corresponding
time delay, as it passes by a massive object, have been experimentally
observed.  Hence experimental confirmation of the bending of the
trajectory of a phonon will give further credence to our view that the
phenomena of bending of the trajectory of a massless particle (and the
corresponding time delay) in curved geometries, are features that are
generic to Lorentzian geometry and independent of the dynamics of the
system under consideration. In fact it would liberate these phenomena
from the almost stereotypical association with gravitating systems
alone.  For a given de Broglie wavelength of a phonon, there is a
restriction on its impact parameter. If the impact parameter of a
phonon is smaller than its de Broglie wavelength then the phonon
behaves quantum mechanically and hence both the concepts of trajectory
and impact parameter are meaningless. On the other hand, classical
analysis of phonon propagation can be done, when the de Broglie
wavelength is very very small compared to the impact parameter. So, if
we consider the semiclassical scattering of phonon in acoustic
geometry, the minimum value of the impact parameter should be equal to
the de Broglie wavelength of a phonon. Therefore the maximum value of
the angle of bending (up to 2nd order) of the phonon trajectory and
the corresponding time delay (up to 1st order) are given by,
\begin{equation}
\delta\phi\Big{|}_{max}=\frac{3~\pi}{4}\left(\frac{r_{ergo}}{\lambda_{ph}}
\right)^{2}
\label{phimax} 
\end{equation}

\begin{equation}
\delta t\Big{|}_{max}=\frac{2 B}{c^{2}}\tan^{-1}
\left(\sqrt{\left(\frac{r_{V}}{\lambda_{ph}}\right)^{2}-1}~\right)
\label{tmax} 
\end{equation}
where $\lambda_{ph}$ is the wave-length of phonon.
\bigskip
\par Since vortex motion with a sink has not yet been realized in the
laboratory, it is not possible to give a correct estimate of the angle
of bending and the corresponding time delay, but it is certainly
possible if we consider perfect vortex flow(A=0). For a perfect vortex
flow equations~(\ref{phimax})and (\ref{tmax})~are modified to,
\begin{equation}
\delta\phi\Big{|}_{max}=\frac{3~\pi}{4}\left(\frac{r_{ergo}}{\lambda_{ph}}
\right)^{2}
\label{phimax0} 
\end{equation}

\begin{equation}
\delta t\Big{|}_{max}=\frac{2 B}{c^{2}}\tan^{-1}\left(\sqrt{\left(\frac{r_{V}}{\lambda_{ph}}\right)^{2}-1}~\right)
\label{tmax0} 
\end{equation}
where $r_{ergo}= \frac{B}{c}$.

Instead of a single phonon if we consider two phonons propagating in
the same direction but on opposite sides of the vortex, they will be
seen to converge to a point. Thus a vortex in an inviscid, barotropic
fluid acts like a convergent lens and the corresponding focal length(up
to 2nd order) is given by,
\begin{equation}
f_{min}=\frac{2}{3\pi}~\frac{\lambda_{ph}^{3}}{r_{ergo}^{2}}
\end{equation}
The circulation for a perfect vortex flow $(
\vec{v}=B/r~\hat{\phi})$ is defined as,
\begin{equation}
\Gamma=\oint \vec{v}.d\vec{x}=2\pi B
\end{equation}
If we consider a singly quantized vortex$(B = \hbar/m)$, then the
circulation of the vortex is,$\Gamma=\frac{h}{m}$. \\
As discussed in \cite{visser} there are two cases where we see perfect 
vortex motion, 
\begin{enumerate}
\item Superfluid Helium,${}^{4}He$,
\item Bose-Einstein condensation in atomic gases.
\begin{enumerate}
\item Bose-Einstein condensation of ${}^{23}Na$,
\item Bose-Einstein condensation of ${}^{87}Rb$,
\end{enumerate}
\end{enumerate}
(1)~$\bf{Superfluid~Helium,{}^{4}He :}$~In case of superfluid
helium\cite{feyn}, \cite{richard}, \cite{huang},\cite{don},
\cite{mand} the energy spectrum of the quasiparticle, as shown by
Feynman, is,
\begin{equation}
\epsilon(k) =\frac{\hbar^{2}~k^{2}}{2~m~S(k)} 
\end{equation}
where $S(k)$ is the structure factor of the liquid which can be
determined from neutron scattering. For small k, $\epsilon(k)$ is
linear and for$~k=1.92 \mathrm{{~\AA}^{-1}} $, S(k) has a maximum .
This corresponds to a minimum in $\epsilon(k)$. The excitations around
that minimum are called rotons and those corresponding to the linear
part of the spectrum are called phonons. In case of superfluid helium
the roton minimum occurs at wave length $\lambda=3.27\mathrm{\AA}$ and
the spectrum is phononic in the wavelength range roughly three
times of it. At temperature very close to $~0^{o}K~$, the speed of sound
in helium is 240 m/s.
So for a singly  quantized vortex in liquid Helium,
\begin{itemize} 
\item 

Wavelength of phonon  $(\lambda_{ph})= 9.81 \mathrm{\AA}  $.
\item 

Radius of the ergo sphere,$(r_{ergo})=0.655 \mathrm{\AA}  $.
\item 

Maximum angle of bending of phonon,$~\delta\phi\Big{|}_{max}=0.6^\circ$
\item 

Minimum focal length,$f\Big{|}_{min}=467 \mathrm{\AA}$
\item 

Maximum  time delay is, $\delta t\Big{|}_{max}=0.857 ps $.
\end{itemize} 
\bigskip 
(2)~$~\bf{Bose-Einstien~Condensate:}$ ~In case of a dilute
Bose-Einstein condensate \cite{smith,legg,parkins,fetter,kurn,mewes,anderson} the energy spectrum of the quasiparticle
excitation is of the Bogoliubov type and is given by,
\begin{equation}
\epsilon_{k}=\sqrt{(\epsilon^{0}_{k})^{2}~+~2~n~U_{0}~\epsilon^{0}_{k}}
\label{bec}
\end{equation}
where~$\epsilon^{0}_{k}=\frac{\hbar^{2}k^{2}}{2~m}$,
~$~U_{0}=\frac{4\pi\hbar^{2}a}{m}~$is the strength of the effective
interaction between the atoms, $n$ is the density of atoms and m is
the mass of the atoms. This spectrum is not similar to that of
superfluid helium. In superfluid helium the quasiparticle excitations
exhibit roton minimum because of strong short range correlation
between the helium atoms, but due to the diluteness of the atomic
vapor quasiparticle excitation spectrum does not exhibit roton
minimum.  The Bogoliubov spectrum is linear in the low wavenumber
limit and quadratic in the large wavenumber limit. The transition from
the linear to quadratic spectrum occurs when $\epsilon^{0}_{k}\sim n
U_{0}$. The reduced wavelength of the phonon corresponding to this
transition point is called coherence length$(\xi)$. For wavelengths
less than the coherent length of the quasiparticle the spectrum is
phononic. Since the velocity of sound is $~c=\sqrt{n U_{0}/m}~$, the
coherent length$(\xi)=\hbar/\sqrt{2}mc$.  For a singly quantized
vortex $r_{ergo}=\hbar/mc $, so
$2\pi\xi=\lambda = \sqrt{2}\pi r_{ergo} $.\\
\\
(a)~$\bf{BEC~of~Sodium,{}^{23}Na:}$ In experiments conducted by the
MIT group, a BEC of sodium atoms was produced with the following
parameters:
\begin{itemize} 
\item 
Number density of Sodium atoms,$(n)=4\times 10^{20}~m^{-3}$.
\item
Speed of sound,$(c)=10.4~mm/s$ .
\item
Coherence length $(\xi)=0.187~\mu m $.
\end{itemize}
\par Using the above data the following are calculated to be:
\begin{itemize}
\item 
Wavelength of phonon $(\lambda_{ph})=1.172~\mu m $.
\item 
Radius of the ergo sphere,$(r_{ergo})=0.263~\mu m $.
\item 
Maximum angle of bending of phonon,$~\delta\phi\Big{|}_{max}=6.84^\circ$
\item 
Minimum focal length,$f\Big{|}_{min}=4.91~\mu m $
\item 
Maximum  time delay is, $\delta t\Big{|}_{max}=80~\mu s $.
\end{itemize} 

\bigskip
\noindent (b)~$\bf{BEC~of~Rubidium,{}^{87}Rb:}$
We also have from the JILA experiment using Rubidium:\\
\begin{itemize} 
\item 
Number density of Rubidium atoms,$(n)=2.6\times 10^{20}~m^{-3}$.
\item  
Speed of sound,$(c)=3.0~mm/s$ .
\item 
Coherence length $(\xi)=0.165~\mu m $.
\end{itemize}

Using this data-set the calculated values are found to be:
\begin{itemize}
\item 
Wavelength of phonon  $(\lambda_{ph})=1.037~\mu m $.
\item 
Radius of the ergo sphere,$(r_{ergo})=0.233~\mu m $.
\item 
Maximum angle of bending of phonon,$\delta\phi\Big{|}_{max}=6.84^\circ$
\item 
Minimum focal length,$f\Big{|}_{min}=4.37~\mu m $
\item 
Maximum  time delay is, $\delta t\Big{|}_{max}= 0.237~ms $.
\end{itemize} 
Looking at the order of magnitude of the observable, it is evident that
BEC in atomic gases are more suitable than liquid helium for
experimental verification of the bending of a phonon trajectory and the
corresponding time delay. Even though the order of magnitude of the
observables for the BEC is small, given  the recent advances in
technology, they are detectable.
\section{Discussion}
\label{discuss}
In acoustic analogue models of gravity, propagation of a sound wave in
an inhomogeneous flow of barotropic inviscid fluid is equivalent to that
of a minimally coupled massless scalar field in curved acoustic geometry
which is Lorentzian in signature. In the eikonal limit we can treat a
sound wave as a ray and also consider only the propagation of a single
phonon instead of a sound ray in the acoustic geometry. In this limit
the trajectory of a phonon is a null geodesic in the acoustic geometry.
When such a phonon travels through an inhomogeneous flow of a barotropic
inviscid fluid, its trajectory is bent. From the perspective of acoustic
geometry this is the bending of the trajectory of a massless particle
due to the curvature of the acoustic geometry.  Due to this bending of
the trajectory, a phonon has to travel along a path which is longer than
the path that it would have taken had the fluid been at rest. In the
short wavelength limit, the leading terms in the expansion of the
bending angle in powers of the velocity parameters, are quadratic. This
result is consistent with Fischer and Visser's\cite{Fischer} result for an
irrotational vortex without a sink. In the same limit the expansion for
the time delay in powers of the velocity parameters ($A$ and $B$) to
first order, is independent of $A$. This phenomenon in acoustic geometry
is similar to the light bending effect and the corresponding time delay
in a physical gravitational system when light passes by a massive
object. 

Instead of a single phonon if we consider two phonons propagating in the same
direction but on opposite sides of the vortex, they will be seen to converge to
a point. Thus a vortex in an inviscid, barotropic fluid acts like a convergent
lens. Both the angle of bending and the time delay in the case of a BEC are
significantly larger than those for liquid helium. In this respect BEC's in
atomic gases are more suitable than liquid helium for experimental verification
of these effects. Since this acoustic analog model of gravity is based on the
assumption that when a sound wave passes through a flowing fluid, the
linearized perturbations of dynamical quantities are stable, demonstrating the
stability of the vortex under such perturbations is more important an issue
than concerns about the order of the magnitude of the observable. This is one
of the directions in which this work can be extended in future.\\
\\
\begin{center}
  ACKNOWLEDGEMENT
  \\
  I would like to thank my advisor Prof. Parthasarathi Majumdar for
  several useful discussions. I am also grateful to Saha Institute of
  Nuclear Physics for their local hospitality, where a part of this
  work was carried out.
\end{center}

\end{document}